\documentstyle[12pt]{article}
\begin{document}
{\hbox to\hsize{August 1992 \hfill UMDHEP 93-020}}\par
{\hbox to \hsize{   \hfill LMU-12/92, SISSA 149/92/EP}}\par

\begin{center}
\vglue .3in
{\LARGE \bf {Planck Scale Effects on the\\
\vglue .4truecm
 Majoron }}\\[.4in]

{\bf E.Kh. Akhmedov$^{1,2}$, Z.G. Berezhiani$^{3,4}$, R.N. Mohapatra$^5$\\
and G. Senjanovi\'{c}$^6$}\\
\vglue .4truecm
{\it ${}^{1}$Scuola Internazionale Superiore di Studi Avanzati\\
I-34014 Trieste, Italy}\\
{\it ${}^{2}$Kurchatov Institute of Atomic Energy, Moscow 123182, Russia}\\
{\it ${}^{3}$Sektion Physik der Universit$\ddot{a}$t M$\ddot{u}$nchen\\
D-8000 M$\ddot{u}$nchen, Germany\\
${}^{4}$Institute of Physics, Georgian Academy of Sciences\\
380077 Tbilisi, Georgia}\\
{\it ${}^{5}$Department of Physics, University of Maryland\\
College Park, Maryland  20742, U.S.A.}\\
{\it ${}^{6}$International Centre for Theoretical Physics\\
I-34014 Trieste, Italy}\\[.2in]

{\bf {Abstract}}\\[.1in]
\end{center}
\begin{quotation}
The hypothesis that non-perturbative gravitational effects lead to
explicit breaking of global symmetries is considered in the
context of Majoron models.
We find that the nonvanishing Majoron mass generated by these
effects can overclose the universe unless the massive Majoron
is unstable. The cosmological mass density constraints
can then be satisfied only if $V_{BL} < 10$ TeV, where $V_{BL}$
is the scale of $B-L$ symmetry breaking.

\end{quotation}
\newpage

\noindent {\bf {1.}}	The idea that neutrinos may be massive Majorana
particles has been extensively discussed in literature.  If this possibility
is realized in nature, it would imply that $B-L$ symmetry of the standard
model is broken by new interactions beyond the standard model, and will
provide a  window to this new physics.  It was proposed in
1980$^{[1]}$ that $B-L$ may be global
symmetry of nature spontaneously broken by the vacuum state.  In this case,
there appears a massless spin zero Nambu-Goldstone boson in the particle
spectrum called the Majoron, which significantly alters our thinking about
the early universe and the role of neutrinos in its evolution.

\vspace{2mm}

The original idea of the Majoron was realized in an extension of the
standard model where the $B-L$ symmetry was spontaneously broken by the
vacuum expectation value of an electroweak singlet complex scalar boson. The
associated Majoron is called the singlet Majoron model$^{[1]}$. Subsequently,
this idea was realized in models where the $B-L$ symmetry was spontaneously
broken by the vev's of iso-triplet$^{[2]}$ and iso-doublet$^{[3]}$ scalar
bosons leading to the so-called triplet and doublet Majoron models,
respectively.  The measurement of the $Z$-width at LEP has ruled out the
triplet and doublet Majoron models.  At present, the only viable Majoron
model is the singlet model or mixed Majoron$^{[4]}$ models, where tiny
admixtures of non-singlet components appear in combination with a dominantly
singlet Majoron.

\vspace{2mm}

An important question in these models is the scale $V_{BL}$ at which the $B-L$
symmetry breaks.  Experimentally, this scale will manifest itself in the
Majorana mass of the neutrino.  For instance, in the see-saw models for
neutrino masses, the neutrino mass scales inversely with $V_{BL}$.  In the
singlet Majoron model or its generalizations, $V_{BL}$ is an arbitrary
parameter.  In this letter, we argue that if non-perturbative gravitational
effects explicitly  break all global symmetries,
as has recently been postulated$^{[5,6,7]}$,
the scale $V_{BL}$ of the singlet Majoron model (or its extensions) will
have an upper limit of about a few TeV, or so.\\[.1in]

\noindent {\bf {2.}}	An important ingredient of the singlet Majoron model
is a complex Higgs singlet $\sigma$ with $B-L=-2$, which couples to the
right-handed neutrino $\nu_R$, and acquires a non-zero vev: $<\sigma> =
V_{BL}/\sqrt{2}$.  One can write:

$$\sigma ~ = ~ {1\over \sqrt{2}} ~~ (V_{BL} + \rho)~e^{i {\chi / V_{BL}}}
\eqno(1) $$

In the absence of gravitational effects, $\chi$ is massless and denotes the
singlet Majoron field.  In the early universe, for $T >> V_{BL}, ~<\sigma>$
vanishes and the Majoron field acquires a mass.  In the mixed Majoron models,
the true Majoron field has only $\chi$ as the dominant piece, and the
discussion given below applies.

\vspace{2mm}

Once the gravitational effects are turned on, $U(1)_{B-L}$ breaking terms
will be induced.  We assume that these terms respect the local symmetries
of the theory and vanish as $M_{P\ell} \longrightarrow \infty$. One can
then write these induced terms that involve only the singlet field $\sigma$
and the standard model Higgs doublet $\varphi$, in which case the lowest
dimension terms are $d=5$:

$$V_{P\ell} ~ = ~ V_1 (\sigma) ~ + ~ V_2 (\varphi, \sigma) \eqno(2) $$
where
$$V_1 (\sigma) ~ = ~ \lambda_1 ~ {\sigma^5 \over M_{P\ell}} ~ + ~ \lambda_2
{}~ {{\sigma^* \sigma^4} \over {M_{P\ell}}} ~ + ~ \lambda_3 {{\sigma^{*^2}
\sigma^3} \over {M_{P\ell}}} + h.c. \eqno(3a) $$
$$V_2(\varphi,\sigma)= \beta_1 ~ {{(\varphi^{\dagger}\varphi)^2 \sigma}
\over {M_{P\ell}}} ~ + ~ \beta_2 ~ {{(\varphi^{\dagger}\varphi)\sigma^2
\sigma^*}\over{M_{P\ell}}} ~ + ~ \beta_3~{{(\varphi^{\dagger}\varphi)
\sigma^3}\over{M_{P\ell}}}~+~h.c. \eqno(3b) $$
We will further assume that $\lambda$ and $\beta$ are not small.  $V_{P\ell}$
will now induce non-vanishing mass for the Majoron field $\chi$.  A massive
Majoron will affect the evolution of the universe, unless its mass and
lifetime satisfy certain constraints.  This, in turn, leads to restrictions on
the value of the $B-L$ breaking scale $V_{BL}$.  In order to discuss this
question, we consider two complementary ranges of the parameter $V_{BL}$:

\noindent (A)  $V_{BL} < V$ and (B)  $V_{BL} > V$, where $V =
(\sqrt{2} G_F)^{-{1 \over 2}} \simeq 246$ GeV.  In case (A), the Majoron
mass is dominated by the $\beta_1$-term in eq. (3),
whereas in case (B), the dominant terms are
from $\lambda_1, \lambda_2$, and $\lambda_3$ couplings.  In case (A), the
Majoron mass is estimated to be (estimation valid for $m_{\chi}\leq V_{BL}$

$$m_\chi ~ \simeq ~ \beta_1^{1 \over 2} ~ \left( {V \over V_{BL}}
\right)^{1 \over 2} ~ {\rm keV} \eqno(4) $$
\noindent On the other hand, in case (B), we get

$$m_\chi ~ \simeq ~ \left( {25 \over 2} \lambda_1 ~ + ~ {9 \over 2} \lambda_2
{}~ + ~ {1 \over 2} \lambda_3 \right)^{1 \over 2} ~ \left( {V_{BL} \over V}
\right)^{3 \over 2} ~ {\rm keV} \eqno(5) $$

Obviously, the mass acquired by the Majoron is minimal when $V_{BL}=V$.
We see that if $\lambda$'s and $\beta$'s are of order 0.1 to 1, the
Majoron mass is larger than a keV.  Therefore, unless its lifetime is
appropriately constrained, the Majoron population of the universe will
overclose it. As a rough estimate of the situation, we see from the
desired constraint

$$n_\chi m_\chi ~ < ~ \rho_{crit} \eqno(6) $$

\noindent that, if $n_\chi \simeq n_\gamma$, then for $V = V_{BL}$,
we must have $\lambda, \beta < 10^{-2}$.  As $V_{BL}$ becomes
significantly different
from $V$, the situation is worse and the Majoron must be unstable.  Of
course, if it happens that $\chi$ decouples sufficiently early, leading
to $n_\chi << n_\gamma$, constraints on $\chi$ may be weaker.  To study
the implications of a massive Majoron for cosmology, we therefore need
to know the following:\\
\noindent i)   When does the Majoron go out of equilibrium?\\
\noindent ii)  How does it decay?

\vspace{2mm}

Both these questions are tied to the nature of interactions that involve
the $\sigma$-field.  Typically, the following interaction terms dictate
the answers to the above questions:

$$L_Y ~ = h_{ab} ~ \bar{\psi}_{aL} ~ \varphi ~ \nu_{iR} ~ + ~ f_{ab} ~
\nu_{aR}^T ~ C^{-1} ~ \nu_{bR} ~ \sigma ~ + ~ h.c. \eqno(7) $$
\vspace{2mm}
\noindent
Here $\psi_{aL}$ denotes the leptonic doublet, $\nu_{aR}$ is the
right-handed neutrino field and $a,~b,$... stands for the generation index.
Eq. (7) leads to the see-saw formula for neutrino masses when we substitute
$< \varphi^o > = {V \over \sqrt{2}}$ and $< \sigma > = {{V_{BL}} \over
{\sqrt{2}}}$, which yields:

$$h ~ \approx ~ {{m_\nu^D} \over {V}} \sqrt{2} ~ \approx ~
{{\sqrt{2m_\nu m_{\nu_R}}} \over {V}} \eqno(8) $$
(Here and below, we consider only the heaviest of the neutrinos,
denoted by $\nu$, since only its effects are the most significant).
Since in our framework the Majorons are massive, the left-handed neutrinos
are stable and cosmological mass density constraint requires that $m_\nu
< 25$ eV. This implies that

$$h ~ \leq ~ 10^{-6} \left( {m_{\nu_R} \over {\rm GeV}} \right)^{1 \over 2}
\eqno(9) $$

It is then easy to see that $\nu_R$ is in equilibrium with the left-handed
electrons and neutrinos via the interaction $\psi_L + \phi \longrightarrow
\nu_R + W_L$ (which has an interaction rate $\Gamma$ given by $\Gamma \approx
{{g^2 h^2} \over {16 \pi}} T$), unless $m_\nu \leq 10^{-4}$ eV.

\vspace{2mm}

This is in equilibrium for $10^5 m_{\nu_R} \geq T > m_{\nu_R}$.  Since the
$\sigma$-field is in equilibrium with the right-handed neutrinos via the
Yukawa coupling $f$ (if the latter is not unnaturally small),
they are in equilibrium with the rest of the universe.
As the universe cools below $T = m_{\nu_R}$, the right-handed neutrinos
disappear.  The dominant interactions of $\sigma$ is via its induced coupling
to $\nu_L$'s.  This coupling has a magnitude

$$f_{\nu_L \nu_L \sigma} ~ \simeq ~ \left( {m_\nu \over V_{BL}} \right)
\eqno(10) $$
For $V_{BL}$ in the 1 GeV to 100 GeV range, $f_{\nu_L \nu_L \sigma} \simeq
2 \times 10^{-8} - 2 \times 10^{-10}$.  Therefore, for $m_{\nu_R}>10$
GeV or so, $\sigma$'s go out of equilibrium at $T = m_{\nu_R}$.  Their
present density is then given by

$$r_{\chi}={n_\chi (T_o) \over n_\gamma (T_o)} ~ = ~
 {g_* (T_o) \over g_* (T_D)} ~
\approx ~ {1 \over 10} - {1 \over 20} \eqno(11) $$

In eq. (11), $g_* (T)$ denotes the effective number of light particle species
at temperature $T$; $T_D$ and $T_o$ denote the decoupling temperature for the
right-handed neutrinos and the present temperature of the universe
respectively.

\vspace{2mm}

Eq. (6) then implies that $m_\chi < 2$ keV.  Using eq. (4) and (5), one
finds for all $\lambda$'s and $\beta$'s of the order one, that neither case
(A) nor case (B) are acceptable unless the Majoron decays.  On the other hand,
if $\lambda_i < 10^{-1}$, then $V_{BL}\simeq V$ is  acceptable  for the
$B-L$ symmetry breaking scale.\\[.1in]

\noindent {\bf {3.}} Now let us look at the constraints on $V_{BL}$ arising
when the Majoron is unstable. According to eq. (10), the dominant decay mode,
$\chi\rightarrow \nu\nu$, is provided by the heaviest of the ordinary
neutrinos. Then the lifetime is given by

$$\tau_{\chi}=8\pi\left(V_{BL}\over{m_\nu}\right)^2 {m_\chi}^{-1} \eqno(12)$$
The constraint to be satisfied in this case is$^{[8]}$

$$r_{\chi}m_\chi \left( {\tau_\chi \over \tau_U} \right)^{1 \over 2} ~<~25
({\Omega}_o h^2)~{\rm eV} \eqno(13) $$

\noindent where  $\tau_U$
is the age of the universe. This leads to

$$r_{\chi}\left(m_{\chi}\over 1~{\rm keV}\right)\left(\frac{\tau_{\chi}}
{1~{\rm sec}}\right)^{1/2}\leq 10^{7}\,\Omega_{o}h^{3\over 2} \eqno(14)$$

\vspace{2mm}

A more stringent constraint comes from the requirements of galaxy formation
in the universe. The recent COBE measurements of cosmic background
anisotropy$^{[9]}$ implies a very small magnitude for initial density
fluctuations: $\delta\rho/\rho\leq {10}^{-4}$. This, in turn, requires a
sufficiently long matter dominated epoch for the linear growth of
$\delta\rho/\rho$ to form the observed large scale structure of the universe.
Therefore, the relativistic decay products of $\chi$ must be redshifted enough
so as to have a matter dominated universe by the era, $t_{eq}$, where one has
the equality of matter and radiation density $^{[10]}$ ($t_{eq}=0.8\times
10^{3}\left(\Omega h^2\right)^{-2}$ yr). This leads to

$$r_{\chi}~n_{\gamma}(t_{eq})~m_\chi \cdot\left( {\tau_\chi \over
t_{eq}} \right)^{1 \over 2} ~ < ~ \rho_{M}(t_{eq}) \eqno(15) $$
where $n_{\gamma}(t_{eq})=(1+z_{eq})^3\times422$ cm${}^{-3}$ is the
photon number density at that epoch and  $\rho_M(t_{eq})=(1+z_{eq})^3
\times 10.5(\Omega_{o}h^2)~{\rm keV}\cdot{\rm cm^{-3}}$ is the energy
density of non-relativistic (presumably cold dark) matter.
This leads to the condition:

$$r_{\chi}\left( \frac{m_{\chi}}{1~{\rm keV}}\right)\left(\frac
{\tau_{\chi}}{1~{\rm sec}}\right )^{1 \over 2} < 4\times 10^3 \eqno(16)$$
(Notice that this condition does not depend on the value of
$\Omega_{o}h^2$). Assuming $r_{\chi}=0.1$, this can be rewritten as\\
$$m_{\chi}\left(V_{BL}\over{V}\right)^2\leq 10^6\left(m_\nu\over{25~
{\rm eV}}\right)^2~{\rm keV}  \eqno(17)$$
Using eq. (5) and assuming the numerical factors in it to be of order
one, we get:\\
$$V_{BL} <\left(m_\nu\over{25~{\rm eV}}\right)^{4/7}\times{10~{\rm TeV}}
\eqno(18)$$
If we use eq. (4) instead of eq. (5), the result is an upper limit on
$V_{BL}$ bigger than the electro-weak scale (where eq. (4) does not hold).
We thus conclude that the limit on the scale of $B-L$ breaking due to Planck
scale effects inducing $B-L$ breaking terms of dimension 5, is given by eq.
(18) and is in the TeV range. The limit actually does not depend on the
particular model of the singlet Majoron and applies to the original
see-saw model$^{[1]}$ as well as the modified versions $^{[4]}$.

\vspace{2mm}

An important implication of our result is that the neutrino masses must have
a lower bound if they arise via the see-saw formula.  The precise value of the
lower bound, of course, depends on the magnitude of the Dirac mass of the
neutrino.  In particular, since the $B-L$ scale is so low, one of the neutrinos
could have a mass in the eV range which in turn implies that a few
per cent of the  dark matter  of the universe being hot. This may indeed
be indicated by the recent COBE data.\\[.1in]

\noindent {\bf {4.}}	Our considerations could easily be extended to a
wider class of theories where the leading order $B-L$ violating term may be
bigger than 5.  For instance, if the leading order term is of the form
${\sigma^{4+n} \over M_{P\ell}^n}$, then

$$V_{BL}<(10)^{\frac{10}{n+6}}\left({M_{P\ell}\over{{\rm GeV}}}\right)^{n
\over n+6}~\left({m_\nu \over 25~{\rm eV}}\right)^{4\over n+6}~{\rm GeV}
\eqno(19)$$

\noindent Thus, for higher $n$, bounds on $V_{BL}$ are less stringent.  It
is possible to construct gauge models where the Planck scale effects can be
postponed to higher dimensional terms.  We do not discuss specific models in
this paper.

\vspace{2mm}

In conclusion, we point out that Planck scale effects can impose interesting
upper limits on the scale of $B-L$ breaking in Majoron models.  These
considerations can be easily extended to familon and other models with
spontaneously broken global symmetries, where the pseudo-Goldstone bosons
couple to neutrinos.

\vspace{4mm}

{\large \center{Acknowledgements}}\\[.1in]

We would like to thank G. Fiorentini, E. Bellotti and the organizers of the
Gran Sasso Neutrino Workshop where this work was initiated and A. Dolgov,
A. Smirnov and M. Vysotsky for useful discussions.  After this work was
completed, we were informed that similar ideas are also under consideration
by K.S. Babu, I.Z. Rothstein, and D. Seckel (to appear as Bartol preprint).
The work of R.N.M. is supported by a grant from the National Science
Foundation and that of Z.B. is supported by the Alexander von Humboldt
Foundation.


\vspace{6mm}

\begin{thebibliography}{99}

\bibitem{chikashige} Y. Chikashige, R.N. Mohapatra and R.D. Peccei,
{\it {Phys. Lett.}} {\underline{98B}}, 265 (1981).
\bibitem{gelmini} G. Gelmini and M. Roncadelli, {\it {Phys. Lett.}}
{\underline {99B}}, 411 (1981); H. Georgi, S.L. Glashow and S. Nussinov,
{\it {Nucl.Phys.}} {\underline {B193}}, 297 (1981).
\bibitem{aulakh} C. Aulakh and R.N. Mohapatra, {\it {Phys. Lett.}}
{\underline {119B}}, 136 (1983); S. Bertolini and A. Santamaria,
{\it {Nucl.Phys.}} {\underline {B310}}, 714 (1988).
\bibitem{berez}K.S. Babu and R.N. Mohapatra, {\it {Phys.
Rev.Lett.}} {\underline {67}}, 1498 (1991); K. Choi and A. Santamaria,
San Diego Preprint (1991); A. Joshipura, PRL Preprint (1991);
Z. Berezhiani, A.Yu. Smirnov and J.W.F. Valle, Preprint FTUV/92-20 (1992).
\bibitem{giddings} S. Giddings and A. Strominger, {\it Nucl. Phys.}
{\underline{B307}}, 854 (1988); S. Coleman, {\it Nucl. Phys.}
{\underline {B310}}, 643 (1988); S.J. Rey, {\it Phys. Rev.} {\underline
{D39}}, 3185 (1989); B. Carter, in {\it General Relativity}, "An Einstein
Centenary Survey", ed. by S. Hawking and W. Israel, Cambridge University
Press, (1979).
\bibitem{barbieri} R. Barbieri, J. Ellis and M.K. Gaillard, {\it {Phys.
Lett.}} {\underline {90B}}, 249 (1980); E. Akhmedov, Z. Berezhiani and
G. Senjanovi\'{c}, preprint IC/92/79, (1992).
\bibitem{holman} R. Holman, S. Hsu, T. Kephart,
E. Kolb, R. Watkins and L. Widrow, {\it Phys. Lett.} {\underline {282B}},
132 (1992); M. Kamionkowski and J. March-Russel,
{\it Phys. Lett.} {\underline {282B}}, 137 (1992);
 S. Barr and D. Seckel, {\it Phys. Rev.} {\underline{D46}},
539 (1992); D. Grasso, M. Lusignoli and M. Roncadelli, {\it Phys. Lett.}
{\underline {288B}}, 140 (1992); S. Ghigna, M. Lusignoli and M. Roncadelli,
{\it Phys. Lett.} {\underline {283B}}, 278 (1992).
\bibitem{dicus} D. Dicus, E. Kolb, V. Teplitz and R. Wagoner,
{\it Phys. Rev.} {\underline {D18}}, 1829 (1978).
\bibitem{smoot} G.F.Smoot {\it et. al.}, {\it {Ap. J. Lett.}} (in press)
(1992).
\bibitem{steigman} G. Steigman and M. Turner, {\it Nucl. Phys.} {\underline
{B253}}, 375 (1985).
\end{thebibliography}
\end{document}